\documentclass[english]{achemso}
\usepackage[LGR,T1]{fontenc}
\usepackage[latin9]{inputenc}
\usepackage{amsmath}
\usepackage{graphicx}
\usepackage[numbers]{natbib}

\makeatletter

\title{First-Passage Time Fluctuation Theorem and Thermodynamic Bound in
Cooperative Biomolecular Networks}

\author{D. Evan Piephoff}

\author{Jianshu Cao}

\email{jianshu@mit.edu}

\affiliation{Department of Chemistry, Massachusetts Institute of Technology, Cambridge,
Massachusetts 02139, United States}

\DeclareRobustCommand{\greektext}{%
  \fontencoding{LGR}\selectfont\def\encodingdefault{LGR}}
\DeclareRobustCommand{\textgreek}[1]{\leavevmode{\greektext #1}}
\ProvideTextCommand{\~}{LGR}[1]{\char126#1}

\makeatother

\usepackage{babel}
\begin{document}
\singlespacing
\begin{abstract}
Using a pathway analysis technique, a dynamic fluctuation relation
is derived for a kinetically cooperative biomolecular machine (e.g.,
an enzyme or a motor protein). This new relation indicates that, in
the absence of hidden current, a fluctuation theorem can be established
for the first-passage time of the observable process, and we show
that this dramatic reduction is a general feature applicable to a
wide variety of cooperative networks. This first-passage time fluctuation
theorem can be experimentally tested, with its violation serving as
a signature of hidden detailed balance breaking. Importantly, we obtain
a compact exact expression for the integrated correction to this fluctuation
theorem, as well as the general single-loop form, revealing a thermodynamic
bound on the kinetic branching ratio (a measure of directionality
defined as the forward-to-backward observable process probability
ratio). These results provide detailed insight into the rich connections
between dynamic measurements and the underlying nonequilibrium thermodynamics
for cooperative biomolecular machines.
\end{abstract}

Advances in spectroscopic techniques have afforded the ability to
observe real-time trajectories of biomolecules at the single-molecule
level.\citep{Moerner_2003,Park2007} These time traces provide insights
into microscopic mechanisms that are usually unavailable from ensemble-averaged
measurements.\citep{English2005} Unique to single-molecule experiments
is the measurement of probability distribution functions (PDFs) of
the waiting times between detectable molecular events, such as the
first-passage time (i.e., the process completion time) PDF. Biomolecular
machines, such as motor proteins\citep{Svoboda1994,Keller2000,Sivak_2022}
or enzymes,\citep{Seifert_2012,Lu_Z_2023} consume energy and dissipate
heat to perform a particular cellular function (e.g., cargo transport,
catalysis, etc.). As such, they operate out of equilibrium, often
in a nonequilibrium steady-state (NESS). In the nonequilibrium setting,
fluctuation relations\citep{Jarzynski1997,Crooks1999} demonstrate
the properties of the PDF for a certain thermodynamic quantity, such
as work or entropy production.\citep{Seifert_2012} Recently, a time-based
fluctuation theorem was derived\citep{Roldan_2015,Neri_2017} for
the first-passage time of entropy production, i.e., the time necessary
to produce a certain amount of entropy. This fluctuation theorem implies
equivalence between the normalized forward and backward entropy production
first-passage time PDFs. On the basis of chemical kinetics, an example
of this equivalence---referred to as the generalized Haldane relation---has
been derived elsewhere\citep{Qian_2006,Ge_2008} for the forward and
backward first-passage time PDFs for a generalized, one-dimensional
(1D) enzymatic chain reaction.

In this kinetic chain, all first-passage trajectories produce the
same amount of entropy. However, for biomolecular machines in which
the driven, observable process is coupled to a hidden process in a
kinetically cooperative\citep{Fersht_1985,Cao2011,Wu2012,Miller_2015,Piephoff_2017,Mu_2021}
fashion, this is not necessarily the case since such trajectories
may begin and end in different underlying states; therefore, this
fluctuation theorem is no longer directly applicable to the first-passage
time of the observable process (even though it is valid for the appropriate
internal distributions). In fact, single-molecule experiments have
revealed the existence of slow, hidden conformational fluctuations
on time scales commensurate to those for the observable process;\citep{English2005}
however, many theoretical treatments have neglected their role due
to the complexity of the calculations involved. These challenges motivate
some important questions for this type of system: (i) under what circumstances
can such a first-passage time fluctuation theorem be established,
(ii) when it cannot (which is experimentally verifiable), what does
this reveal about the hidden dynamics, and (iii) can the general deviation
from this relation be quantified?

To address these questions, we consider the canonical model for such
a system: a kinetic scheme for conformation-modulated single-enzyme
catalysis (a type of continuous-time Markov process) under NESS conditions,
which is relevant to \textgreek{b}-galactosidase\citep{English2005}
and human glucokinase.\citep{Miller_2015,Mu_2021} A novel, efficient
pathway analysis technique (that reduces to the transition rate matrix
approach but is more general)\citep{Cao2008,Piephoff_2018} is adapted
to this model and used to perform kinetic evaluations, allowing us
to attain surprising and concise results from complex calculations.
The key results of this Letter are represented in eqs \ref{eq:fpt-fluc-thm-dev},
\ref{eq:tau-fluc-thm-deltamu}, and \ref{eq:integr-fpt-fluc-thm-dev-gen-form}--\ref{eq:br-ratio-bound}.
We derive a new dynamic fluctuation relation, which indicates that,
in the absence of hidden current, a fluctuation theorem---and by
extension, the generalized Haldane relation---can be established
for the observable process first-passage time; we demonstrate that
this dramatic reduction is a general feature applicable to a wide
variety of cooperative networks. This first-passage time fluctuation
theorem (as well as the Haldane relation) can be tested experimentally,
and its violation serves as a signature of hidden detailed balance
breaking. Furthermore, we obtain a compact exact expression for the
integrated correction to this fluctuation theorem, as well as its
general single-loop form, revealing a thermodynamic bound on the kinetic
branching ratio (a measure of directionality defined as the forward-to-backward
probability ratio for the observable process).

\begin{figure}
\includegraphics[width=0.85\columnwidth]{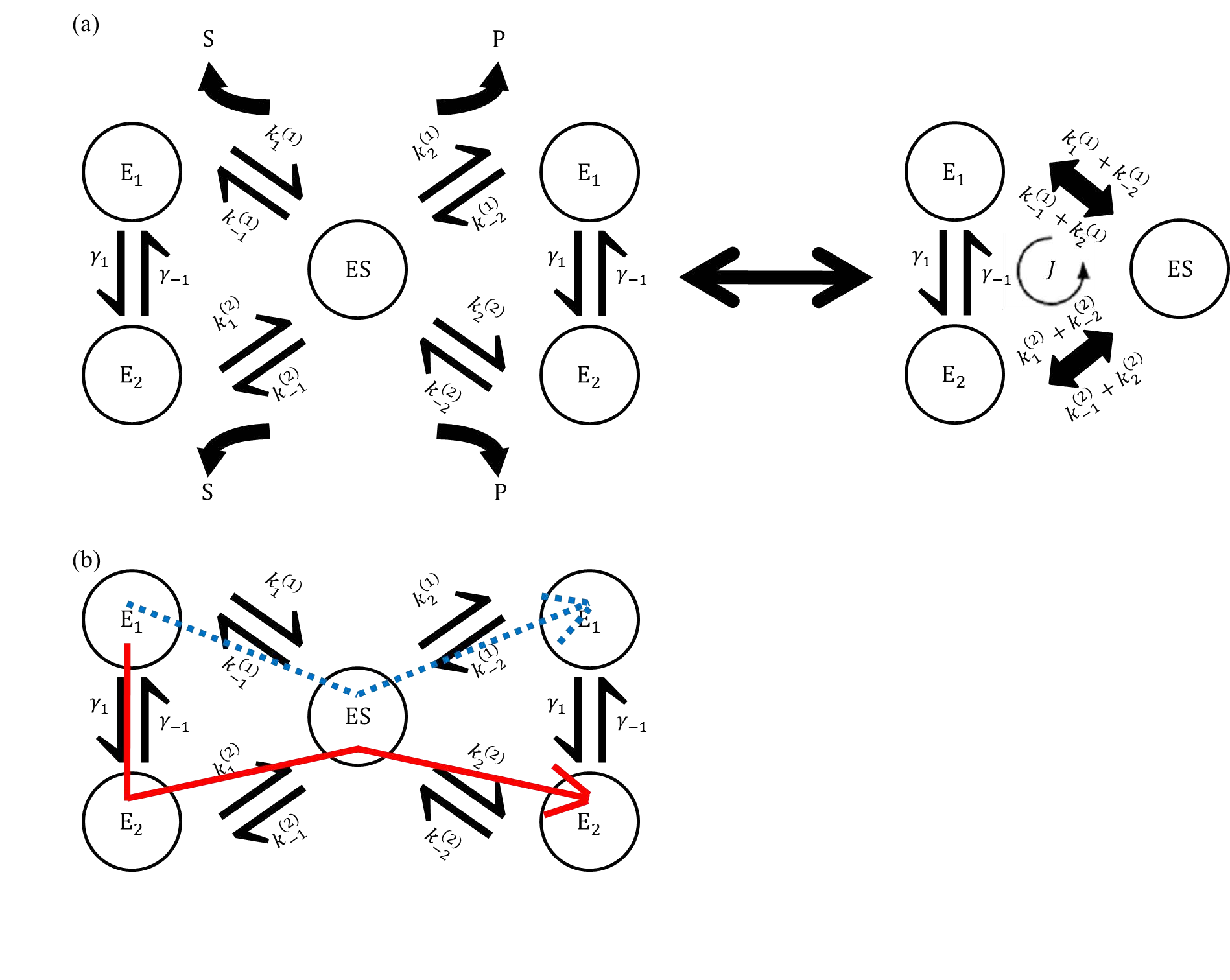}

\caption{\label{fig:enz-model}(a) Minimal model for conformation-modulated
enzyme turnover with kinetic cooperativity (a type of continuous-time
Markov process) under NESS conditions. A single enzyme reversibly
catalyzes the conversion of a substrate (S) to a product (P). The
free enzyme (i.e., the initial state manifold, with states E\protect\textsubscript{1}
and E\protect\textsubscript{2}) can reversibly bind the substrate
(with rates $\left\{ k_{\pm1}^{\left(l\right)}\right\} $), resulting
in the formation of the substrate-bound enzymatic complex (state ES),
which can then reversibly undergo product formation (with rates $\left\{ k_{\pm2}^{\left(l\right)}\right\} $).
Substrate is consumed to form product in the forward observable process,
and product is consumed to form substrate in the backward one. The
single enzyme is embedded in a solution (that serves as a heat bath)
of substrate and product, such that the substrate and product concentrations
(and chemical potentials) remain fixed, and the nonlinear substrate
binding and reverse product formation kinetic transitions are treated
as pseudolinear. The reaction is cooperatively coupled to a hidden
process, as the unbound enzyme undergoes slow conformational interconversion
(based on simple thermal changes, with rates $\left\{ \gamma_{\pm1}\right\} $).
The transition rates are constrained by the local detailed balance
conditions in eqs \ref{eq:ldb-enz-k} and \ref{eq:ldb-enz-conf}.
The right-hand side is a representation of the scheme wherein the
two steps in each reaction pathway are folded onto each other, resulting
in a conformational loop with a corresponding population current $J$.
Such conformation-modulated enzymatic models have experimental relevance
to \textgreek{b}-galactosidase\citep{English2005} and human glucokinase\citep{Miller_2015,Mu_2021}
(see Figure \ref{fig:complex-networks}c,d). (b) Depiction of two
first-passage trajectories for the model in (a), each producing a
different amount of entropy, with the dotted, blue one starting and
ending in the same underlying state, and the solid, red one doing
so in different states.}
\end{figure}

\begin{figure}
\includegraphics[width=1\columnwidth]{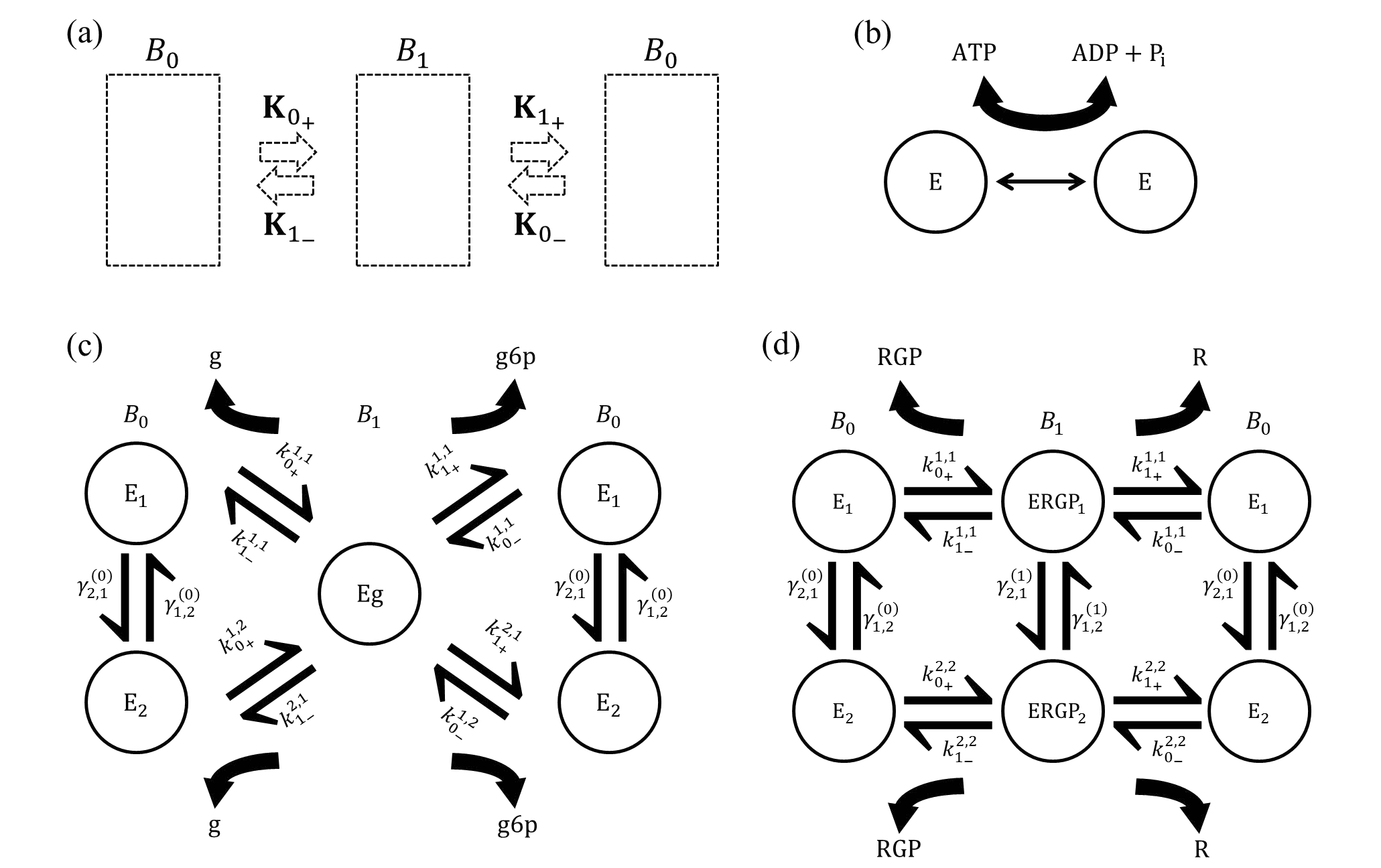}

\caption{\label{fig:complex-networks}(a) Generalization of the model in Figure
\ref{fig:enz-model}a to a generic biomolecular machine with kinetic
cooperativity. The machine undergoes a driven, observable, cyclic
process that is cooperatively coupled to a hidden process with dynamics
occurring within the state manifolds, $\left\{ B_{m}\right\} $. Each
state in $B_{0}$ and $B_{1}$ has an observable process transition
entering and exiting it; otherwise, the internal topologies of the
manifolds are arbitrary, as are their connectivities. Transitions
between manifolds are designated here as $\left\{ \mathbf{K}_{m_{\pm}}\right\} $.
(b) Overall reaction for kinesin catalyzing the hydrolysis of ATP
to ADP and P\protect\textsubscript{i}, which powers its $8$-$\mathrm{nm}$
stepping motion along a microtubule. Under typical experimental conditions,
this process can be described as a single step occurring on the microsecond
time scale, devoid of any substeps.\citep{Carter2005} (c)--(d) Examples
of experimentally relevant underlying schemes corresponding to the
generic kinetic model in (a). The rates of transitions between the
discrete states are represented by $\left\{ k_{m_{\pm}}^{k,l}\right\} $
and $\left\{ \gamma_{k,l}^{\left(m\right)}\right\} $. In (c), the
enzymatic model in Figure \ref{fig:enz-model}a is applied to human
glucokinase, which catalyzes the phosphorylation of glucose (g) to
glucose-6-phosphate (g6p) in glycolysis.\citep{Miller_2015} For an
extended discussion of a three-state model sufficiently capturing
the reaction kinetics for this system, we refer readers to previous
work in ref \citenum{Mu_2021}. In (d), this enzymatic model is adapted
to a scheme with multiple intermediate states that describes \textgreek{b}-galactosidase,
which catalyzes the hydrolysis of resorufin-\textgreek{b}-\textsc{d}-galactopyranoside
(RGP) to resorufin (R).\citep{English2005} In the Supporting Information,
for each of these three specific schemes, values of the experimentally
relevant parameters and conditions are provided, and our results are
examined.}
\end{figure}

Examples of molecular machines\citep{Gingrich2017} include single
enzymes catalyzing the conversion of a substrate to a product,\citep{Lu_2014,Chou2024}
as well as molecular motors transporting cargo.\citep{Sivak_2022}
We begin by considering a minimal model for a kinetically cooperative
biomolecular machine. Figure \ref{fig:enz-model}a depicts a three-state
kinetic scheme for an enzymatic reaction with conformational interconversion
(i.e., the canonical such model). Similar schemes have been employed
previously for different purposes,\citep{Piephoff_2018,Mu_2021} and
such conformation-modulated enzymatic models are experimentally relevant
to \textgreek{b}-galactosidase\citep{English2005} and human glucokinase\citep{Miller_2015,Mu_2021}
turnover (represented in Figure \ref{fig:complex-networks}c,d). Here,
kinetic cooperativity refers to deviation from Michaelis--Menten-like
kinetics due to the presence of population flux through the interconverting
initial states that arises from nonequilibrium conformational dynamics.
This effect has been experimentally observed, including in human glucokinase
catalysis,\citep{Miller_2015} which---as shown in the Supporting
Information---involves a notable conformational current.

We define $\tau_{\pm}$ as the forward/backward first-passage time
for the observable process (i.e., the turnover time), which corresponds
to the time necessary to complete an iteration of the forward/backward
process, while avoiding the completion of the backward/forward one,
which dictates the measurement protocol. An individual trajectory
corresponding to such an iteration is referred to as a forward/backward
first-passage trajectory. The work applied along such a trajectory,
$\pm w$, is referred to as the forward/backward first-passage work,
with $\Delta s^{\mathrm{tot}}=w/T$ for temperature $T$; that is,
we define $\Delta s^{\mathrm{tot}}$ as the entropy production associated
with the first-passage work. Implicit in our analysis is the incorporation
of the entropic contribution of the solution into the dissipated heat.
Accordingly, $w$ corresponds to chemical work, which is defined as
the negative of the free energy change of the solution resulting from
a reaction with stoichiometrically differing total chemical potentials
between the reactants and products (see ref \citenum{Seifert_2012}
for further details). The unnormalized PDF of $\tau_{\pm}$ is represented
as $P_{\pm}\left(\tau_{\pm}\right)$. When all forward/backward first-passage
trajectories produce entropy $\pm\Delta s^{\mathrm{tot}}$, we can
write a fluctuation theorem for $\tau_{\pm}$ that relates the ratio
$P_{+}\left(t\right)/P_{-}\left(t\right)$ exponentially to $\Delta s^{\mathrm{tot}}$,\citep{Qian_2006,Roldan_2015,Neri_2017}
which we refer to as the first-passage time fluctuation theorem (see
eq \ref{eq:tau-fluc-thm-deltamu}). However, in a kinetically cooperative
biomolecular machine, the entropy produced along first-passage trajectories
can vary, since they may begin and end in different underlying states
(as shown in Figure \ref{fig:enz-model}b), so this expression is
no longer directly applicable to $P_{\pm}\left(t\right)$ (even though
it is valid for the appropriate internal distributions). In order
to explore such fluctuation relations in cooperative biomolecular
networks, we will first examine $P_{\pm}\left(t\right)$ for the minimal
model in Figure \ref{fig:enz-model}a.

Let $\mu^{\mathrm{S}}$ represent the chemical potential of the substrate,
and $\mu^{\mathrm{P}}$ represent that of the product. The reaction
process is driven by the difference in chemical potential between
the substrate and product (i.e., the chemical affinity), $-\Delta\mu\equiv\mu^{\mathrm{S}}-\mu^{\mathrm{P}}$,
that is,

\begin{equation}
w=-\Delta\mu=T\Delta s^{\mathrm{tot}}\label{eq:w-deltamu}
\end{equation}
Local detailed balance\citep{Seifert_2012} constrains the transition
rates here as

\begin{equation}
\frac{k_{1}^{\left(1\right)}k_{2}^{\left(1\right)}}{k_{-1}^{\left(1\right)}k_{-2}^{\left(1\right)}}=\frac{k_{1}^{\left(2\right)}k_{2}^{\left(2\right)}}{k_{-1}^{\left(2\right)}k_{-2}^{\left(2\right)}}=\exp\left[-\frac{\Delta\mu}{k_{\mathrm{B}}T}\right]\label{eq:ldb-enz-k}
\end{equation}

\begin{equation}
\frac{\gamma_{1}k_{1}^{\left(2\right)}k_{-1}^{\left(1\right)}}{\gamma_{-1}k_{-1}^{\left(2\right)}k_{1}^{\left(1\right)}}=1\label{eq:ldb-enz-conf}
\end{equation}
where $k_{\mathrm{B}}$ is the Boltzmann constant. Therefore, the
kinetics are described by eight independent rates. It is noted that
eq \ref{eq:ldb-enz-conf} represents the local detailed balance condition
for the closed substrate loop. A similar condition can be written
for the product loop, $\gamma_{1}k_{-2}^{\left(2\right)}k_{2}^{\left(1\right)}/\left(\gamma_{-1}k_{2}^{\left(2\right)}k_{-2}^{\left(1\right)}\right)=1$,
which is implied by eqs \ref{eq:ldb-enz-k} and \ref{eq:ldb-enz-conf};
however, only two independent constraints can be imposed here.

Now, we evaluate $P_{\pm}\left(t\right)$ for the enzymatic model
in Figure \ref{fig:enz-model}a using a novel pathway analysis technique
(that reduces to the transition rate matrix approach but is more general)\citep{Cao2008,Piephoff_2018}
and examine the corresponding first-passage time fluctuation theorem.
Our kinetic approach is based upon the decomposition of a scheme into
generic structures that have corresponding waiting time distribution
functions. We write such functions in terms of self-consistent pathway
solutions and concatenate them using a tensor framework to efficiently
construct $P_{\pm}\left(t\right)$, taking all transitions as first-order
kinetic rate processes (see the Supporting Information for further
details). An event-averaged initial condition is employed that corresponds
to the NESS condition for the measurement protocol described above;
all results presented herein depend upon this initial distribution.
In our model, the addition of the hidden loop and the effect of the
kinetic reversibility significantly complicate the calculations for
$P_{\pm}\left(t\right)$. However, because we simplify the problem
by breaking the connectivity of the scheme down to transitions between
state manifolds and examine only waiting time distribution functions
that correspond to paths cyclic about the initial state manifold,
we are able to use lower-dimensional, dense matrices to attain $P_{\pm}\left(t\right)$
in a way that avoids superfluous intermediate calculations. We note,
though, that our results can also be obtained using the rate matrix
approach.

In the Laplace domain, where the Laplace transform of a function $h\left(t\right)$
is $\check{h}\left(z\right)=\int_{0}^{\infty}\mathrm{d}te^{-zt}h\left(t\right)$,
we can write the fluctuation relation (derivations in the Supporting
Information)

\begin{equation}
\frac{\check{P}_{+}\left(z\right)}{\check{P}_{-}\left(z\right)}-\exp\left[\frac{\Delta s^{\mathrm{tot}}}{k_{\mathrm{B}}}\right]=\check{\alpha}\left(z\right)J\label{eq:fpt-fluc-thm-dev}
\end{equation}
Here, $\check{\alpha}\left(z\right)$ is a complicated rational expression
that is, in general, finite for zero $J$, where $J$ (depicted on
the right-hand side of Figure \ref{fig:enz-model}a) is the hidden
(conformational) population current, normalized by the total hidden
rate $\gamma=\gamma_{1}+\gamma_{-1}$, i.e., $J=\gamma^{-1}\left(\rho_{\mathrm{E}_{1}}^{\mathrm{s}}\gamma_{1}-\rho_{\mathrm{E}_{2}}^{\mathrm{s}}\gamma_{-1}\right)$,
with the stationary population of state $\mathrm{E}_{l}$ represented
as $\rho_{\mathrm{E}_{l}}^{\mathrm{s}}$. It is noted that $\check{\alpha}\left(z\right)$
is not independent of $J$. Also, the second term on the left-hand
side of eq \ref{eq:fpt-fluc-thm-dev} is specified by eqs \ref{eq:w-deltamu}
and \ref{eq:ldb-enz-k}. In addition, $J\propto\left[\gamma_{1}u_{1}^{\left(2\right)}u_{-1}^{\left(1\right)}/\left(\gamma_{-1}u_{-1}^{\left(2\right)}u_{1}^{\left(1\right)}\right)-1\right]$
here, with $u_{\pm1}^{\left(l\right)}=k_{\pm1}^{\left(l\right)}+k_{\mp2}^{\left(l\right)}$.
The hidden (conformational) detailed balance condition, under which
$J=0$, can then be expressed as

\begin{equation}
\frac{\gamma_{1}u_{1}^{\left(2\right)}u_{-1}^{\left(1\right)}}{\gamma_{-1}u_{-1}^{\left(2\right)}u_{1}^{\left(1\right)}}=1\label{eq:hdb-enz}
\end{equation}
From eqs \ref{eq:ldb-enz-k}, \ref{eq:ldb-enz-conf}, and \ref{eq:hdb-enz},
we see that local detailed balance alone is insufficient in satisfying
hidden detailed balance here. If the two steps in each reaction pathway
were folded onto each other (as shown on the right-hand side of Figure
\ref{fig:enz-model}a), then the satisfaction of hidden detailed balance
would correspond to the probability of traversing the resulting loop
being directionally invariant. Thus, eq \ref{eq:hdb-enz} can be thought
of as the cycle balance condition for this folded scheme. It is not
a condition corresponding purely to transitions within an isolated
coarse-grained state, as it depends, in part, upon the connectivity
between the initial state manifold and the rest of the network. Because
the conformational changes here are induced thermally rather than
by external driving, there is no additional conformational affinity
that directly corresponds to $J$. Instead, the conformational current
is an indirect result of the overall chemical affinity ($-\Delta\mu$),
and is an emergent feature of the cooperative kinetics.

When hidden detailed balance (eq \ref{eq:hdb-enz}) is satisfied,
we can establish the first-passage time fluctuation theorem,

\begin{equation}
\frac{P_{+}\left(t\right)}{P_{-}\left(t\right)}=\exp\left[\frac{\Delta s^{\mathrm{tot}}}{k_{\mathrm{B}}}\right]\label{eq:tau-fluc-thm-deltamu}
\end{equation}
recovering the form obtained previously for a generalized, 1D kinetic
chain\citep{Qian_2006} (with the time dependence on the left-hand
side dividing out). That is, for zero $J$, eq \ref{eq:fpt-fluc-thm-dev}
dramatically reduces to a fluctuation theorem equivalent to the one
derived by Roldán and Neri et al.\citep{Roldan_2015,Neri_2017} for
the first-passage time of entropy production, as all forward/backward
first-passage trajectories here now produce entropy $\pm\Delta s^{\mathrm{tot}}$.
The deviation from the first-passage time fluctuation theorem is thus
governed by the hidden current. This result is surprising because,
under zero $J$, $P_{\pm}\left(t\right)$ does not generally reduce
to the 1D chain form, but the ratio $P_{+}\left(t\right)/P_{-}\left(t\right)$
does, indicating a unique reduction in how the forward and backward
observable processes relate to one another. The normalized PDF corresponding
to $P_{\pm}\left(t\right)$ is given by $\phi_{\pm}\left(t\right)=P_{\pm}\left(t\right)/p_{\pm}$,
with forward/backward observable process probability $p_{\pm}=\int_{0}^{\infty}\mathrm{d}tP_{\pm}\left(t\right)$,
where $p_{+}+p_{-}=1$. Equation \ref{eq:tau-fluc-thm-deltamu} implies
the symmetry relation

\begin{equation}
\phi_{+}\left(t\right)=\phi_{-}\left(t\right)\label{eq:hald-rel-1-1}
\end{equation}
which is referred to as the generalized Haldane relation and has been
derived for a 1D enzymatic chain reaction.\citep{Qian_2006,Ge_2008}
Similarly, it was shown that for a generalized, 1D kinetic chain involving
a motor protein, the mean forward and backward first-passage times
are equal\citep{Kolomeisky2005} (as implied by eq \ref{eq:hald-rel-1-1});
this was also demonstrated experimentally for kinesin translocation.\citep{Carter2005}
The PDF $\phi_{\pm}\left(t\right)$ can be readily measured experimentally;
therefore, the generalized Haldane relation can be tested, and its
violation---which implies a violation of the first-passage time fluctuation
theorem (eq \ref{eq:tau-fluc-thm-deltamu})---serves as a signature
of hidden detailed balance breaking. Similar results connecting waiting
time statistics to hidden currents have been reported elsewhere,\citep{Ptaszynski2018,Parrondo2019,Dunkel2021,Seifert2022,Roldan2022}
but the present analysis shows how such a signature of hidden detailed
balance breaking follows naturally from the novel fluctuation relation
in eq \ref{eq:fpt-fluc-thm-dev}. It is noted that $P_{\pm}\left(t\right)$
(as well as $w$) can also be measured (see the Supporting Information
for further details); thus, the first-passage time fluctuation theorem
can be directly tested itself. In the Supporting Information, we examine
our results in the context of three experimental systems: kinesin
stepping, human glucokinase catalysis, and \textgreek{b}-galactosidase
turnover (depicted in Figure \ref{fig:complex-networks}b--d).

For multiple hidden loops due to the presence of more than two states
in the initial state manifold $B_{0}$ (see Figure S2 in the Supporting
Information for an example), when all but one of the resulting cooperative
hidden currents vanish, the basic form of eq \ref{eq:fpt-fluc-thm-dev}
holds for the single unbalanced current (see the Supporting Information
for further details); thus, when that current also vanishes, the first-passage
time fluctuation theorem (eq \ref{eq:tau-fluc-thm-deltamu}) and generalized
Haldane relation (eq \ref{eq:hald-rel-1-1}) are recovered (the same
is true for a cooperative scheme with multiple states in the intermediate
state manifold $B_{1}$, such as the one in Figure \ref{fig:complex-networks}d).
Equations \ref{eq:fpt-fluc-thm-dev}, \ref{eq:tau-fluc-thm-deltamu},
and \ref{eq:hald-rel-1-1} are therefore quite general, with this
signature of hidden detailed balance breaking being applicable to
a wide variety of cooperative networks (i.e., those corresponding
to the generic kinetic model in Figure \ref{fig:complex-networks}a).
The generality of this signature is further confirmed by our subsequent
work,\citep{stoch-therm-arxiv-v1-acs} which uses a stochastic thermodynamics
framework to understand hidden detailed balance breaking in a generic,
cooperative biomolecular machine. Additionally, we note that, while
$P_{\pm}\left(t\right)$ is typically a lengthy expression, the form
of eq \ref{eq:fpt-fluc-thm-dev} here is quite simple and general.

\begin{figure}
\includegraphics[width=1\columnwidth]{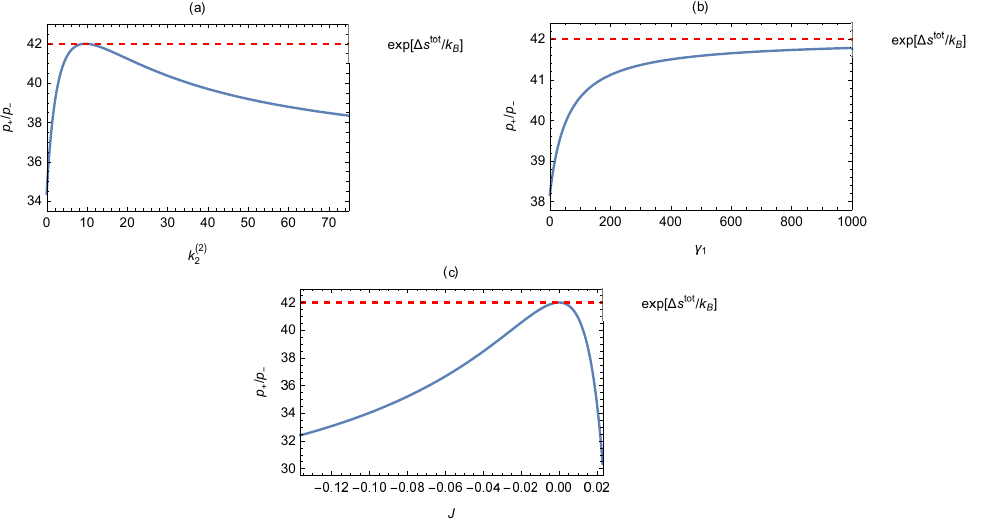}

\caption{\label{fig:br-ratio-plots}Plots of $p_{+}/p_{-}$ (solid, blue curves)
against $k_{2}^{\left(2\right)}$ (a), $\gamma_{1}$ (b), and $J$
(c) for the model in Figure \ref{fig:enz-model}a under the following
conditions: $k_{1}^{\left(1\right)}=70$, $k_{-1}^{\left(1\right)}=5$,
$k_{-1}^{\left(2\right)}=16$, $k_{2}^{\left(1\right)}=3$, and $k_{-2}^{\left(1\right)}=1$
(the units of the rates here are arbitrary), such that $\exp\left[\Delta s^{\mathrm{tot}}/k_{\mathrm{B}}\right]=42$.
In (a) and (b), $k_{-2}^{\left(2\right)}=0.2$, with $\gamma_{1}=40$
in (a), and $k_{2}^{\left(2\right)}=40$ in (b). In (c), for plotting
purposes, the local detailed balance constraints (eqs \ref{eq:ldb-enz-k}
and \ref{eq:ldb-enz-conf}) are substituted in with $k_{2}^{\left(2\right)}$
and $\gamma_{-1}$ {[}instead of $k_{1}^{\left(2\right)}$ and $\gamma_{-1}$,
as is the case in (a) and (b){]}, and $J$ is substituted in with
$k_{-2}^{\left(2\right)}$, with its range restricted to positive
values of $k_{2}^{\left(2\right)}$ and $k_{-2}^{\left(2\right)}$
($\gamma_{-1}$ is independent of $J$ here); $k_{1}^{\left(2\right)}=40$
and $\gamma_{1}=0.2$. The behavior of $p_{+}/p_{-}$ is monotonic
in (b), whereas in (a) and (c), a turning point is observed when the
hidden current vanishes, saturating the bound in eq \ref{eq:br-ratio-bound}
{[}it is saturated in the asymptotic limit in (b){]}. In each case,
it is seen that this bound is obeyed, with $\exp\left[\Delta s^{\mathrm{tot}}/k_{\mathrm{B}}\right]$
depicted by a dashed, red line.}
\end{figure}

Continuing with the model in Figure \ref{fig:enz-model}a, we now
analyze the deviation from hidden detailed balance. The local detailed
balance constraints (eqs \ref{eq:ldb-enz-k} and \ref{eq:ldb-enz-conf})
are substituted in with $k_{1}^{\left(2\right)}$ and $\gamma_{-1}$,
such that $\exp\left[\Delta s^{\mathrm{tot}}/k_{\mathrm{B}}\right]=k_{1}^{\left(1\right)}k_{2}^{\left(1\right)}/\left(k_{-1}^{\left(1\right)}k_{-2}^{\left(1\right)}\right)$
and $J\propto\left[k_{2}^{\left(1\right)}/k_{-1}^{\left(1\right)}-k_{2}^{\left(2\right)}/k_{-1}^{\left(2\right)}\right]$.
We note that this particular choice of rates for substitution is arbitrary,
as the basic forms of eqs \ref{eq:fpt-fluc-thm-dev} and \ref{eq:integr-fpt-fluc-thm-dev-gen-form}
hold, irrespective of how local detailed balance is satisfied (for
instance, an alternative substitution choice is demonstrated in Figure
\ref{fig:br-ratio-plots}c). Evaluating eq \ref{eq:fpt-fluc-thm-dev}
at $z=0$, the integrated correction to the first-passage time fluctuation
theorem can be expressed as (see the Supporting Information)

\begin{equation}
\frac{p_{+}}{p_{-}}-\exp\left[\frac{\Delta s^{\mathrm{tot}}}{k_{\mathrm{B}}}\right]=-\zeta^{\mathrm{eff}}J^{2}\label{eq:integr-fpt-fluc-thm-dev-gen-form}
\end{equation}
where

\begin{equation}
\zeta^{\mathrm{eff}}J^{2}=\frac{\mathcal{D}^{-1}k_{1}^{\left(1\right)}}{k_{-1}^{\left(1\right)}+k_{-1}^{\left(2\right)}}\left(\exp\left[\Delta s^{\mathrm{tot}}/k_{\mathrm{B}}\right]-1\right)\left(k_{2}^{\left(1\right)}k_{-1}^{\left(2\right)}-k_{2}^{\left(2\right)}k_{-1}^{\left(1\right)}\right)^{2}\label{eq:fric-coef-jsq}
\end{equation}
with

\begin{equation}
\begin{split}\mathcal{D} & =\gamma_{1}k_{-1}^{\left(1\right)^{2}}k_{2}^{\left(1\right)}+\gamma_{1}k_{-1}^{\left(1\right)}k_{-1}^{\left(2\right)}k_{2}^{\left(1\right)}+k_{-1}^{\left(1\right)}k_{-1}^{\left(2\right)}k_{1}^{\left(1\right)}k_{2}^{\left(1\right)}+\gamma_{1}k_{-1}^{\left(1\right)}k_{2}^{\left(1\right)^{2}}+k_{-1}^{\left(2\right)}k_{1}^{\left(1\right)}k_{2}^{\left(1\right)^{2}}\\
 & \,\,\,\,\,\,\,+\gamma_{1}k_{-1}^{\left(1\right)^{2}}k_{2}^{\left(2\right)}+\gamma_{1}k_{-1}^{\left(1\right)}k_{-1}^{\left(2\right)}k_{2}^{\left(2\right)}+k_{-1}^{\left(1\right)}k_{-1}^{\left(2\right)}k_{1}^{\left(1\right)}k_{2}^{\left(2\right)}+k_{-1}^{\left(1\right)^{2}}k_{-2}^{\left(1\right)}k_{2}^{\left(2\right)}+k_{-1}^{\left(1\right)}k_{-1}^{\left(2\right)}k_{-2}^{\left(1\right)}k_{2}^{\left(2\right)}\\
 & \,\,\,\,\,\,\,+2\gamma_{1}k_{-1}^{\left(1\right)}k_{2}^{\left(1\right)}k_{2}^{\left(2\right)}+k_{-1}^{\left(1\right)}k_{-2}^{\left(1\right)}k_{2}^{\left(1\right)}k_{2}^{\left(2\right)}+\gamma_{1}k_{-1}^{\left(1\right)}k_{2}^{\left(2\right)^{2}}+k_{-1}^{\left(1\right)}k_{1}^{\left(1\right)}k_{2}^{\left(2\right)^{2}}+k_{-1}^{\left(1\right)}k_{-2}^{\left(1\right)}k_{2}^{\left(2\right)^{2}}
\end{split}
\label{eq:denom-integr}
\end{equation}
Here, $\zeta^{\mathrm{eff}}$ is the effective friction coefficient,
which serves as a measure of (nondissipative) internal ``kinetic
friction'' due to the cooperative hidden dynamics, with $\zeta^{\mathrm{eff}}>0$
($<0$) for $w>0$ ($<0$). It is noted that $\zeta^{\mathrm{eff}}$
is not independent of $J$; for completeness, explicit expressions
for $\zeta^{\mathrm{eff}}$ and $J$ are provided in the Supporting
Information. Also, eqs \ref{eq:integr-fpt-fluc-thm-dev-gen-form}--\ref{eq:denom-integr}
are independent of $k_{-2}^{\left(2\right)}$; that is, $p_{+}/p_{-}$
only depends upon seven independent parameters here, even though $\check{P}_{+}\left(z\right)/\check{P}_{-}\left(z\right)$
depends upon eight. We note that, given the significant complexity
of $\check{P}_{+}\left(z\right)/\check{P}_{-}\left(z\right)$, it
is remarkable that a compact exact expression for $p_{+}/p_{-}$ is
attainable. In addition, as was the case for eq \ref{eq:fpt-fluc-thm-dev},
eq \ref{eq:integr-fpt-fluc-thm-dev-gen-form} holds for an arbitrarily
complex scheme with a single unbalanced cooperative hidden current,
and thus represents the general single-loop form for $p_{+}/p_{-}$.

We define $p_{+}/p_{-}$ as the kinetic branching ratio for the observable
process, which serves as a measure of the deviation from hidden equilibrium
and is obtainable from experimental measurements (see the Supporting
Information for further details). For the model in Figure \ref{fig:enz-model}a,
when $w>0$ (i.e., $\mu^{\mathrm{S}}>\mu^{\mathrm{P}}$), eqs \ref{eq:integr-fpt-fluc-thm-dev-gen-form}--\ref{eq:denom-integr}
imply that

\begin{equation}
\frac{p_{+}}{p_{-}}\leq\exp\left[\frac{\Delta s^{\mathrm{tot}}}{k_{\mathrm{B}}}\right]\label{eq:br-ratio-bound}
\end{equation}
with the direction of the inequality reversed for $w<0$. Thus, the
entropy production associated with the first-passage work (i.e., the
chemical affinity in this case) bounds the kinetic branching ratio,
with the equality recovered under hidden detailed balance (i.e., $J=0$).
In eq \ref{eq:br-ratio-bound}, the left-hand side can be thought
of as a kinetic measure of directionality that accounts for the cooperative
hidden dynamics, whereas the right-hand side corresponds to a thermodynamic
measure of directionality (analogous to the stepping ratio for molecular
motors\citep{Wang_2013}) that does not. The inequality indicates
that, relative to the thermodynamic optimum, the directionality of
the observable process in Figure \ref{fig:enz-model}a is diminished
by what can be envisaged as an internal kinetic-friction-like effect
resulting from the cooperative dynamics. We believe this to be a novel
understanding of how cooperativity can impact directionality. Various
plots of $p_{+}/p_{-}$ demonstrating this bound are shown in Figure
\ref{fig:br-ratio-plots}.

In this Letter, we have explored the first-passage time fluctuation
theorem for a kinetically cooperative biomolecular machine in a NESS.
In this pursuit, the canonical model for such a system, a kinetic
scheme for a conformation-modulated enzymatic reaction (which has
experimental relevance to \textgreek{b}-galactosidase\citep{English2005}
and human glucokinase\citep{Miller_2015,Mu_2021}), has been considered
(see Figure \ref{fig:enz-model}a). We have adapted a novel pathway
analysis framework\citep{Cao2008,Piephoff_2018} (that reduces to
the transition rate matrix formalism) to this model and employed it
to perform kinetic evaluations. A new dynamic fluctuation relation
(eq \ref{eq:fpt-fluc-thm-dev}) has been derived, which reveals that
the deviation from the first-passage time fluctuation theorem is governed
by the hidden current, and in the absence of such current, the first-passage
time fluctuation theorem (eq \ref{eq:tau-fluc-thm-deltamu})---and
by extension, the generalized Haldane relation (eq \ref{eq:hald-rel-1-1})---can
be established. We have demonstrated that this dramatic reduction
is a general result applicable to a wide variety of cooperative networks.
The first-passage time fluctuation theorem (as well as the Haldane
relation) can be tested experimentally, with its violation serving
as a signature of hidden detailed balance breaking. This result is
surprising because, for zero $J$, $P_{\pm}\left(t\right)$ does not,
in general, reduce to the 1D chain form, but the ratio $P_{+}\left(t\right)/P_{-}\left(t\right)$
does, signifying a unique reduction in how the forward and backward
observable processes relate to each other. Additionally, a remarkably
compact exact expression (eqs \ref{eq:integr-fpt-fluc-thm-dev-gen-form}--\ref{eq:denom-integr})
has been obtained for the integrated correction to the first-passage
time fluctuation theorem, as well as the general single-loop form
(eq \ref{eq:integr-fpt-fluc-thm-dev-gen-form}), implying that the
kinetic branching ratio is bounded (eq \ref{eq:br-ratio-bound}) by
the entropy production associated with the first-passage work. This
indicates that the directionality of the observable process here is
diminished by an internal kinetic-friction-like effect resulting from
the cooperative hidden dynamics. In a subsequent study,\citep{stoch-therm-arxiv-v1-acs}
we interpret our signature of hidden detailed balance breaking through
the lens of stochastic thermodynamics, and further demonstrate the
generality of this finding for a broad class of systems.

\section*{Acknowledgments}

We thank Nigel Goldenfeld for helpful discussions. This work was supported
by the Singapore-MIT Alliance for Research and Technology (SMART).
J.C. acknowledges support from the ACS PRF, the NSF DMREF program
(Grant No. CHE-2324300), and the MIT RSC Fund.
\begin{suppinfo}
Adaptation of our pathway analysis framework to the enzymatic model
in Figure 1a, evaluation of the forward and backward first-passage
time distributions and their associated quantities, calculation of
the hidden current, and examination of our results in the context
of experimental systems
\end{suppinfo}

\bibliography{fpt-fluct-thm}

\end{document}